\documentclass[useAMS,usenatbib]{mn2e}
\usepackage{graphicx}
\usepackage{amsmath,amsfonts,amssymb}
\usepackage{epsfig}

\newcommand{\tlife}{$\tau_{\rm life}$}
\newcommand{\tmrg}{$\tau_{\rm mrg}$}
\newcommand{\td}{$\tau_{\rm d}$}
\newcommand{\tc}{$\tau_{\rm c}$}
\newcommand{\npsr}{$N_{\rm psr}$}
\newcommand{\prate}{${\cal P}({\cal R})$}
\newcommand{\grate}{${\cal P}_{\rm g}({\cal R_{\rm g}})$}
\newcommand{\weq}{$W_{\rm eq}$}
\newcommand{\pspin}{P_{\rm s}}

\begin{document}

\title[Galactic NS$-$NS Binary Merger Rate]{Implications of PSR J0737$-$3039B for the Galactic NS$-$NS Binary Merger Rate}
\author[Kim, Perera, \& McLaughlin]
{\parbox[t]{\textwidth}{Chunglee Kim$^{1,2}$\thanks{Email:chunglee.kim0@gmail.com}, Benetge Bhakthi Pranama Perera$^{2,3}$, \& Maura A.\ McLaughlin$^{2}$}\\
\vspace*{3pt} \\
\\ $^{1}$ Astronomy Program, Department of Physics and Astronomy, Seoul National University, 1 Gwanak-ro, Gwanak-gu, Seoul 151-742, Korea
\\ $^{2}$ Department of Physics, West Virginia University, Morgantown, WV 26506, USA
\\ $^{3}$ Jodrell Bank Centre for Astrophysics, School of Physics and Astronomy, University of Manchester, Manchester M13 9PL, UK}

\maketitle

\begin{abstract}
The Double Pulsar (PSR J0737$-$3039) is the only neutron star-neutron star 
(NS$-$NS) binary in which both NSs have been detectable as radio pulsars. 
The Double Pulsar has been assumed to dominate the Galactic 
NS$-$NS binary merger rate ${\cal R}_{\rm g}$ among all known systems, 
solely based on the properties of the first-born, recycled 
pulsar (PSR J0737$-$3039A, or A) with an assumption for 
the beaming correction factor of 6.
In this work, we carefully correct observational biases for the second-born, non-recycled pulsar (PSR J0737$-$0737B, or B)
and estimate the contribution from the Double Pulsar 
on ${\cal R}_{\rm g}$ using constraints available from both A and B.
Observational constraints from the B pulsar favour 
a small beaming correction factor for A ($\sim2$), 
which is consistent with a bipolar model.
Considering known NS$-$NS binaries with the best observational 
constraints, including both A and B,
we obtain ${\cal R}_{\rm g}=21_{-14}^{+28}$ Myr$^{-1}$ 
at 95 per cent confidence from our reference model.
We expect the detection rate of gravitational waves from NS$-$NS inspirals for 
the advanced ground-based gravitational-wave detectors 
is to be $8^{+10}_{-5}$ yr$^{-1}$ at 95 per cent confidence. 
Within several years, gravitational-wave detections relevant to NS$-$NS inspirals 
will provide us useful information to improve pulsar population models.
\end{abstract}


\section{Introduction}

As of today, there are four confirmed neutron star-neutron star (NS$-$NS) binaries\footnote{PSR J1906$+$0746, 
discovered by \citet{J1906discovery}, is the latest known merging NS$-$NS binary 
candidate. However, the nature of its companion is still inconclusive \citep{kasianthesis,ferdman2013} and we do not include this binary in this work.}
in the Galactic plane that will merge within a Hubble time. 
All known NS$-$NS binaries contain at least one radio pulsar that 
is detected by large-scale pulsar surveys: PSRs B1913+16 
\citep{B1913discovery}, B1534+12 \citep{B1534discovery}, the Double Pulsar 
J0737$-$3039 \citep{J0737Adiscovery,J0737Bdiscovery}, and J1756$-$2251 
\citep{J1756discovery}. 
NS$-$NS mergers are one of the most promising sources from which 
detect gravitational waves (GWs) with ground-based interferometers 
\citep[e.g.,][and reference therein]{ligoratespaper2010}. 
By modelling the Galactic disc pulsar population as well as 
selection effects based on observed properties of 
known binaries and survey characteristics,
one can infer the Galactic merger rate estimates 
(${\cal R}_{\rm g}$) 
and GW detection rate (${\cal R}_{\rm det}$) 
for NS$-$NS binaries with ground-based GW detectors 
\citep[e.g.,]{ph91,nps91,cl95,knst,kkl2003,kkl2010,ok2010}.  

The Double Pulsar was discovered in the Parkes high-latitude pulsar survey (Burgay et al. 2003; Lyne et al. 2004)\nocite{J0737Adiscovery, J0737Bdiscovery}. This binary has been assumed to dominate ${\cal R}_{\rm g}$ 
based on the properties of the first-born, recycled pulsar PSR J0737$-$3039A (hereafter A)
due to its large assumed beaming correction factor and short estimated lifetime. 
\citet{kalogera2004} estimated the most likely value of ${\cal R}_{\rm g} \sim 90$ Myr$^{-1}$, 
considering PSRs B1913+16, B1534+12, and the A pulsar. 
Without observational constraints, they assumed A's beaming correction factor to be 6. 
This is an average of the estimated beaming correction factors for PSRs B1913+16 and B1534+12, 
based on polarization measurements.

\citet{ok2010} attempted to calculate the beaming correction factor
for each pulsar found in NS$-$NS and NS$-$white dwarf (NS$-$WD) binaries in the Galactic disc making use of
the latest observations available then.
They estimated A's beaming correction factor adapting the results from 
the polarization measurements \citep{0737Aspin1} and pulse profile analysis \citep{ferdman0737A}. 
They found that, if A is bipolar and an orthogonal rotator ($\alpha \sim 90^{\circ}$), 
its beam must be wide leading to a small beaming correction factor 
($\sim 1.55$ based on their reference model). 
Here, $\alpha$ is the magnetic misalignment angle between the spin and magnetic axes.
If A is unipolar, where the magnetic axis is likely to be aligned with 
the spin axis ($\alpha < 4^{\circ}$), its beam size is unconstrained.
Although they calculated B pulsar's beaming correction factor ($\sim14$) 
motivated by the empirical correlation between a pulsar's beam size and 
spin period, B was still not included in the rate calculation,
due to the lack of information to model this pulsar.
Considering PSRs B1913+16, B1534+12, PSR J0737$-$3039A, J1756$-$5521, and J1906+0746 
with estimated beaming correction factors,
they suggested ${\cal R}_{\rm g}$ is most likely to be $\sim60$ Myr$^{-1}$ 
(the median is $\simeq 89$ Myr$^{-1}$).

The latest pulse profile analysis of A is presented by \citet{ferdman2013}. 
For the pulse widths obtained at $30-50$ per cent of the total pulse height
the corresponding beaming correction factor of A ranges between $\sim3$ and 5.
Low-intensity levels show broader pulse widths, and hence, imply smaller beaming correction factors.
Fitting a two-pole model to pulse widths measured at the 30 per cent of the maximum height,
for example, they obtained $\alpha=90\fdg2^{+11\fdg3}_{-11\fdg4}$, $\rho_{\rm 1}=18\fdg5^{+4\fdg3}_{-0\fdg42}$, 
and $\rho_{\rm 2}=12\fdg1^{+6.1}_{-0.9}$ at 68 per cent confidence.
The variables $\rho_{\rm 1}$ and $\rho_{\rm 2}$ represents half-opening angles 
of the first ($\rho_{\rm 2}$) and second ($\rho_{\rm 1}$) brightest components of the pulse profile, respectively.

PSR J0737$-$3039B (hereafter B) was detectable by the Green Bank Telescope (GBT) 
for almost five years since the discovery \citep{J0737Bdiscovery}. 
The last significant detection made by the GBT was in 2008 March (MJD 54552) 
as reported by \citet{perera2010}. The non-detection of the B pulsar after 2008 
is interpreted as the filled part of B's beam moving completely out of the line of sight 
due to geodetic precession \citep{bo1975}. 
The predicted and measured precession rates of B are 
$5.0347_{\rm -0.0007}^{\rm +0.0007}$ 
and $4.77_{\rm -0.65}^{\rm +0.66}$ deg yr$^{-1}$ at the 68 per cent 
confidence level, respectively \citep{renepsrA}. 
Based on the estimated geodetic precession time-scale for B, 
it will be detectable again in the time window $2013-2035$ 
(Kramer 2010; Perera et al.\ 2010, 2012)\nocite{perera2010,kramerprecession2010,perera2012}. 
The uncertainty in the reappearance time depends on the 
symmetry of the beam function and the exact details of 
the flux gradients across the beam.

The main challenges in modelling B are attributed to its strong 
pulse profile modulations. Pulsar B's secular pulse profile 
change is also evidence of the effects of geodetic precession. 
Moreover, the interaction between A's wind and B's magnetosphere
affects B's pulse profiles over a single orbit. Due to the impact of the wind from 
A \citep{Lyutikov2005}, B was only observable during a fraction of 
its orbital phase, detected as two bright (BP1 and BP2) and two weak 
(WP1 and WP2) phases \citep{J0737Bdiscovery,perera2010}. 
In contrast, A has had an extremely stable pulse profile since its discovery \citep{ferdman0737A}, 
which is consistent with the interpretation that its spin axis is 
likely to be aligned with the orbital angular momentum vector 
(e.g., \citealt{0737formation2006}). 
In this work, we calculate the correction factors to compute the number of 
B-like pulsars in the Galactic disc, adapting results from \citet{perera2012}. 
We are able to better constrain the Galactic NS$-$NS merger rate estimates based on 
the Double Pulsar, by applying the B pulsar's observed properties in addition to those of A.

In Section \ref{sec:empirical}, we briefly describe 
\prate, the probability density function (PDF) of a pulsar binary merger rate estimate 
based on the empirical modelling.
In Sections \ref{sec:model} and \ref{sec:rates}, 
we describe our survey simulations for the B pulsar and
derive \prate~for the Double Pulsar.
Considering PSRs B1913+16 and B1534+12, J0737$-$3039A, and J0737$-$3039B, 
we also calculate the PDF of the Galactic NS$-$NS merger rate (\grate).
We discuss the results in Section \ref{sec:discussion}.

\section{PDF of NS$-$NS Merger Rate Based on a Pulsar Binary}\label{sec:empirical}

Following the same empirical method described in Kim et al.\ (2003, 2010), 
Kalogera et al.\ (2004)\nocite{kkl2003,kalogera2004,kkl2010}, 
and O'Shaughnessy \& Kim (2010)\nocite{ok2010}, we calculate \prate~for 
an NS$-$NS binary population, based on an observed system (e.g., PSR B1913+16), by

\begin{eqnarray} \label{eq:rate}
{\cal P}({\cal R}) &=& {({\tau_{\rm life}}/{N_{\rm pop}})^2}{\cal R}~ \exp{[-({\tau_{\rm life}}/{N_{\rm pop}}){\cal R}]}  \nonumber \\
&\equiv& {C^2} {\cal R}~ \exp[{-C {\cal R}}]~,
\end{eqnarray}
where $\cal R$ is the merger rate estimate, \tlife~is an effective lifetime of the binary and 
$N_{\rm pop}$~is the population size, i.e., the total number of 
pulsars like the observed one in the Galactic disc.
Both \tlife~and $N_{\rm pop}$ depend on the observed properties of the known pulsar and the binary. 
The derivation of equation (\ref{eq:rate}) can be found in section 5.1 in \citet{kkl2003}.
Throughout the paper, we adopt model 6 from \citet{kkl2003}, 
except for the pulsar luminosity function. We describe our assumptions about the pulsar 
luminosity function in Section 3.2

The population size can be obtained by $N _{\rm pop}\equiv$\npsr$ \zeta f_{\rm b,eff}$,
where \npsr~represents the number of detectable pulsars 
like the known pulsar (e.g., the B pulsar) 
among those beaming towards the Earth, given one detection.
$f_{\rm b,eff}$ is the beaming correction factor to take into account a pulsar's finite beam size.
Unlike other pulsars known in NS$-$NS binaries, 
the B pulsar is observable only during certain orbital phases.
We introduce a parameter $\zeta$ to model B-like pulsars, incorporating the observable orbital phases.

We note that equation (\ref{eq:rate}) can be used when an NS$-$NS binary has 
only one detectable pulsar. Although both pulsars in the Double Pulsar have been detected, 
\citet{kalogera2004,kkl2010}, and \citet{ok2010} used equation (\ref{eq:rate}) 
as they considered only the A pulsar in their work.
In \S \ref{sec:rates}, we derive \prate~for the Double Pulsar, 
considering two independent observational constraints from the A and B pulsars.

\section{Pulsar Survey Simulation}\label{sec:model}

We perform Monte Carlo simulations to calculate \npsr~ 
for each known pulsar, modelling a Galactic disc pulsar population 
and pulsar survey sensitivities. 
Below, we summarize our model assumptions and steps to calculate \npsr. 
More details on the modelling can be found in sections 3 and 4 in \citet{kkl2003}, 
including assumptions and related systematic uncertainties in the modelling of the interstellar medium and the pulsar spatial distribution.

We establish a population of pulsars like one of the known pulsars (e.g., the B pulsar), 
by fixing the intrinsic pulse width ($W$) and spin period ($P_{\rm s}$) of model pulsars 
to those of the pulsar.
Each model pulsar's sky location and luminosity are 
randomly sampled from a pulsar luminosity function $p_{\rm L}(L)$ and 
spatial distribution $p_{\rm r}(x,y,z)$. 
All pulsars are assumed to beam towards the Earth.
We assume a Gaussian radial distribution and exponential vertical distribution
that are consistent with the observed pulsars in the Galactic disc 
(see Kim et al.\ 2003\nocite{kkl2003} for further details and the systematic uncertainties
regarding $p_{\rm r}(x,y,z)$ in the rate estimates). 
As for the luminosity distribution, we choose a lognormal distribution
based on the discussion presented in Section \ref{sec:luminosity}.
We emphasize that the empirical rate calculation presented in this work 
as well as other works such as \citet{kalogera2004}
does not involve with observed radio fluxes or distances of known pulsars. 
The only literature that used the observed radio flux (of the A pulsar) 
to infer the Galactic NS$-$NS merger rate is \citet{J0737Adiscovery}.

At a given frequency, the apparent radio flux density of each model pulsar $k$ 
is calculated by $F L_{\rm k}/(x_{\rm k}^{2} + y_{\rm k}^{2} + z_{\rm k}^{2})$, 
where $F$ ($0 < F \le 1$) is a flux degradation factor taken into account the Doppler 
smearing in an orbit and is fixed for the known pulsar. 
When there is no degradation $F=1$. 
The flux degradation factor depends on the known pulsar's spin period, 
pulse width, binary orbital period, eccentricity of the orbit, and the integration time of each survey.
Fast-spinning pulsars in tight orbit normally have small $F$.
For example, the apparent flux density of a pulsar similar to the A pulsar
is only $\sim 15$ per cent of its intrinsic radio flux density
for the Parkes multibeam survey (PMB) with 35-min integration 
time \citep{pmb}. Therefore, we incorporate $F_{\rm PMB}=0.154$ when simulating the PMB survey
for the A-like pulsar population\footnote{\citet{kalogera2004} first incorporated
the flux degradation factor for the A-like pulsars in the survey simulation code. 
For PSRs B1913+16 and B1534+12, we use the estimated $F_{\rm PMB}$ presented in \citet{kkl2003}. 
For other surveys, where the integration time is shorter than that of PMB, the flux degradation effects are not significant.
\citet{ok2010} and this work use the same code they used, 
adding more surveys as mentioned in the text.}.
Due to its longer spin period, however, we can set $F=1$ for all surveys for the B pulsar.

The outcome of the Monte Carlo simulation is $N_{\rm det}$, which
is the number of pulsars brighter than the survey threshold 
among a total of $10^{6}$ realizations. 
Following section 2.1 in \citet{ok2010}, 
we calculate \npsr~by $10^{6}/N_{\rm det}$ for each known pulsar.
This is based on the linear relation
between $N_{\rm det}$ and the number of realization $N$ as described in \citet{kkl2003}. 
See fig.\ 3 and equation (8) in their paper for more details.
Using this relation $N_{\rm det}=sN$, where $s$ is the proportionality constant,
we can write $N/N_{\rm det} = N_{\rm psr}$.

We consider 22 large-scale pulsar surveys in this work, 
including three more surveys to those listed in table 1 in \citet{kkl2003}. 
The two additional surveys, the Parkes multibeam high latitude survey \citep{phsurvey} 
and the mid-latitude drift scan survey with the Arecibo telescope 
\citep{midbo}, are considered in \citet{ok2010} as well. 
The new addition in this work is 
the latest large-scale pulsar survey with the Arecibo L-band Feed Array 
\citep[PALFA;][]{palfaI}.
We adopt the PALFA precursor survey parameters (e.g., 100 MHz bandwidth, 
$40^{\circ} \le l \le 75^{\circ}$ and $168^{\circ} \le l \le 214^{\circ}$ 
with $|b| \le 1^{\circ}$) as described 
in \citet{J1906discovery}. This is the survey that discovered PSR J1906+0746. 
See fig.\ 2 in \citet{palfaI} for the comparison of survey regions between different 
large-scale L-band pulsar surveys including the PALFA precursor survey. 
We assume all survey data are completely processed.

We obtain $N_{\rm pop}$ by applying correction factors to 
compensate for observational biases to $N_{\rm psr}$ for each known pulsar.
We discuss details of important ingredients to modelling the B pulsar in the following subsections.
In Table \ref{tab:ab}, we list the properties of A and B pulsars used in this work.
For PSRs B1913+16 and B1534+12, we use the same parameters listed in table 1 in \citet{ok2010},
but $N_{\rm psr}$ and $N_{\rm pop}$ are recalculated by the latest code including the PALFA survey.

\begin{table} 
\caption{Observational and estimated properties of A and B: Pulsar's spin period ($P_{\rm s}$ in ms), time derivative of spin period ($\dot P_{\rm s}$ in $10^{-18}$ ss$^{-1}$), mass ($M_{\rm psr}$ in solar mass), age estimate ($\tau_{\rm age}$ in Gyr), binary merger time-scale ($\tau_{\rm mgr}$ in Gyr), radio-emission time-scale ($\tau_{\rm d}$ in Gyr), and references. See Section \ref{sec:lifetime} for definitions of the time-scales. 
}
\label{tab:ab}\footnotesize
\begin{center}
\begin{tabular*}{8cm}{lllllllllc}
\hline
PSR & $P_{\rm s}$ & ${\dot{P}}_{\rm s} 10^{-18}$ & $M_{\rm psr}$ & $\tau_{\rm age}$ & $\tau_{\rm mgr}$ & $\tau_{\rm d}$ & Ref.$^{a}$ \\
      & (ms)   & (ss$^{-1}$) & ($M_{\odot}$) & (Gyr)             & (Gyr)  & (Gyr) &   \\
\hline
   A &   22.7 & 1.74$^{b}$   & 1.34        & 0.14              & 0.085  & $>$14 & 1 \\ 
   B & 2770   & 892$^{b} $   & 1.25        & $0.05-0.19^{c}$ & 0.085  & 0.04  & 2 \\
\hline
\end{tabular*}
\end{center}
\medskip
{$^a$References: (1) Burgay et al.\ (2003);\nocite{J0737Adiscovery} (2) Lyne et al.\ (2004).\nocite{J0737Bdiscovery}\\
$^b$Kramer et al.\ (2006). \nocite{kramer2006}\\
$^c$The range of $\tau_{\rm c}$ for the B pulsar is adapted from Lorimer et al.\ (2007). \nocite{J0737age}
}
\end{table}

\subsection{Equivalent Pulse Width} \label{sec:weq}

\begin{figure} 
          \includegraphics[width=8.7cm]{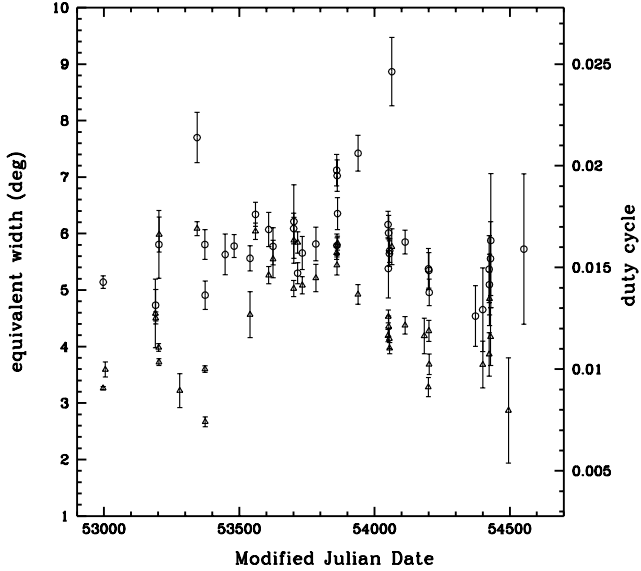} 
          \caption{The measured equivalent pulse width \weq~and duty cycle $\delta$ 
                   obtained from B's pulse profiles in two bright phases, BP1 (triangles) 
                   and BP2 (open circles), respectively. 
                  We use the average value of 
                  \weq$\simeq 4\fdg68$ ($\delta \simeq 0.013$) as our reference.
                   } 
          \label{fig:weq}
        \end{figure}

We calculate B's equivalent pulse width \weq~from 
observed pulse profiles in two bright phases. We define \weq~ as 
the area under the integrated pulse profile divided by 
the maximum peak of the profile. We use \weq~as 
an approximation of an intrinsic pulse width. 
Fig.\ \ref{fig:weq} shows the estimated \weq~from each bright phase 
(triangles for BP1 and open circles for BP2). We find 
that \weq~changes between [$1\fdg9, 9\fdg5$] 
including 1-$\sigma$ errors. The error bars are larger in later 
observations when B became significantly fainter by three to four orders 
of magnitudes. See figs 1 and 2 of \citet{perera2010} for 
actual pulse profiles from BP1 and BP2. 
For a given pulse width, the duty cycle $\delta$ is estimated by 
$W_{\rm eq} {\rm (in~deg)}/360^{\circ}$. 
We use the average duty cycle $\delta \simeq 0.013$ (\weq$\simeq 4\fdg68$) 
as a reference parameter for B in the Monte Carlo simulations.

\subsection{Pulsar Luminosity Distribution}\label{sec:luminosity}

Our reference pulsar luminosity function is described by 
the lognormal distribution with $\left< \log {L} \right>=-1.1$ and 
$\sigma_{\rm \log {L}}=0.9$ \citep{fk2006}, 
motivated by the fact that it does not require a fiducial minimum luminosity.
It is known that both power-law ($p_{\rm L}(L) \propto L^{-2}$) and 
lognormal luminosity distributions are consistent with the current 
pulsar observations, regardless of a pulsar's formation scenario 
(e.g., binaries or singles), location (e.g., disc or globular clusters), 
or spin evolution (e.g., recycling).
\citet{fk2006} studied isolated pulsars in the Galactic disc, 
and suggested that the lognormal distribution 
best fits the observed luminosity distribution of the canonical (i.e., 
non-recycled, young, isolated) pulsar population. Based on 82 isolated 
and binary pulsars found in several globular clusters, \citet{hessels2007gc} 
argued that there is no significant difference in the luminosity distribution 
between isolated and binary pulsars. \citet{hessels2007gc} also 
found that the luminosity distribution of globular cluster pulsars 
can be described by a power-law distribution, which is similar to 
what is proposed by \citet{cc97} based on 22 millisecond pulsars ($P_{\rm s} < 20$ ms) 
found in the Galactic disc. Recently, \citet{bagchi2011} analysed about 
a hundred recycled pulsars found in globular clusters and fit 
the observed pulsar luminosity distribution with power-law and 
lognormal distributions. They concluded that a lognormal distribution 
is a slightly better fit to the observed luminosity distribution 
based on the $\chi^{2}$ and Kolmogorov-Smirnov (K-S) statistics, although both power-law 
and lognormal distributions are, in general, consistent with the observation.

Most of the previous NS$-$NS merger rate estimates used the power-law 
distribution (Kim et al.\ 2003, 2010; Kalogera et al.\ 2004)\nocite{kkl2003,kalogera2004,kkl2010}. The review paper 
on the GW detection rates for compact binary coalescences published 
by the LIGO-Virgo Collaboration \citep{ligoratespaper2010} is also 
based on the merger rate estimates obtained with the power-law distribution.
\citet{ok2010} compared \grate~for NS$-$NS and NS$-$WD binaries obtained 
from the reference power-law distribution ($\propto L^{-2}$) 
with a minimum pseudo-luminosity of 
0.3 mJy kpc$^{2}$ at 1400 Hz used in \citet{kalogera2004} and the 
best-fitting lognormal distributions suggested by \citet{fk2006}. 
They showed that the uncertainty in the peak rate estimate  
due to the choice of the pulsar luminosity distribution 
is less than 10 per cent (see their appendix A for details).
Therefore, assumptions on the pulsar luminosity distribution 
used in this work and previous works are consistent within 
this range of uncertainty.

\begin{figure} 
          \includegraphics[width=8.6cm]{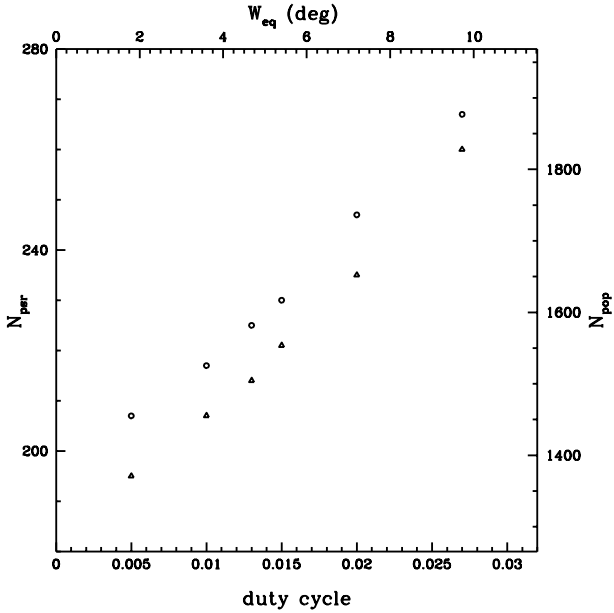}  
          \caption{The number of detectable pulsars like B 
                   among those beaming towards the Earth (\npsr)
                   is shown as a function of duty cycle $\delta$. On the right $y$-axis, 
                   we show the total number of B-like pulsars in the Galactic disc $N_{\rm pop}$.
                   Open circles and triangles are results from power-law and lognormal 
                   luminosity distributions, respectively. Based on our reference 
                   model (lognormal, $\delta \simeq 0.013$), we obtain $N_{\rm pop,B} \sim 1500$.}
          \label{fig:npsr}
      \end{figure}

In Fig.\ \ref{fig:npsr}, we show \npsr~
obtained from the Monte Carlo simulations with different duty cycles for B. 
On the right $y$-axis, we also show the total population size $N_{\rm pop,B}=N_{\rm psr} \zeta f_{\rm b,eff}$,
applying the correction factors calculated in the 
following subsections, $\zeta_{\rm B}=1.9$ and $f_{\rm b,eff,B}=3.7$. 
We consider $\delta=$[0.005, 0.027] as shown in Fig.\ \ref{fig:weq}. 
The \npsr~based on the lognormal distribution (triangles) 
is typically smaller than that of power-law distribution 
(open circles) by a few percent. For our reference duty cycle $\delta \simeq 0.013$, 
we obtain $N_{\rm psr,B} \sim 200$. This implies there are total 
$\sim 1500$ B-like pulsars in the Galactic disc. 
Considering B's pulse profile modulation, 
we expect the number of B-like pulsars in the Galactic disc ranges between $\sim 1300$ and $1800$. 
This can be also read as the total number of the Double Pulsar-like binaries 
in the Galactic disc, based on the properties of the B pulsar.

\subsection{Effective Beaming Correction Factor}\label{sec:fbeff}

A beaming correction factor $f_{\rm b}$ is defined as the 
inverse of the pulsar's beaming fraction, i.e., 
the solid angle swept out by the pulsar's radio beam divided by $4{\rm \pi}$.
The simplest beam model involves only two parameters, 
the half-opening angle of the beam ($\rho$) and magnetic misalignement angle ($\alpha$). 
Assuming all pulsars have two poles with the same beam size of $\rho$, 
we calculate $f_{\rm b}$ as follows 
\begin{equation}\label{eq:fb}
{f_{\rm b}(\alpha,\rho)} = {4{\rm \pi}} {\Bigl[} {2{\rm \pi}}\times 2 \int_{\rm {max}(0, \alpha-\rho)}^{\rm {min}(\alpha+\rho,{\rm \pi}/2)} d\cos\theta~ {\Bigr]}^{-1}~.
\end{equation}
The magnetic misalignment angle of B is estimated based on different assumptions and techniques. 
\citet{perera2012} obtained 
$\alpha=61\fdg0 ^{+7\fdg9}_{-2\fdg4}$ 
at 68 per cent confidence from the pulse profile analysis. 
\citet{renepsrA} estimated $\alpha \sim 70^{\circ}$ 
by fitting a phenomenological model with the eclipse profile of A.
All estimates given in the literature are consistent within 
the 95 per cent confidence level (see table 2 in Perera et al.\ 2010\nocite{perera2010} for a summary). 
Assuming that other parameters needed to describe the beam geometry 
to be relatively constant over time \citep{renepsrA}, we adopt 
the best-fitting value $\alpha=61^{\circ}$ from \citet{perera2012}
as our reference.

\begin{center}
       \begin{figure}
         \includegraphics[width=8.5cm]{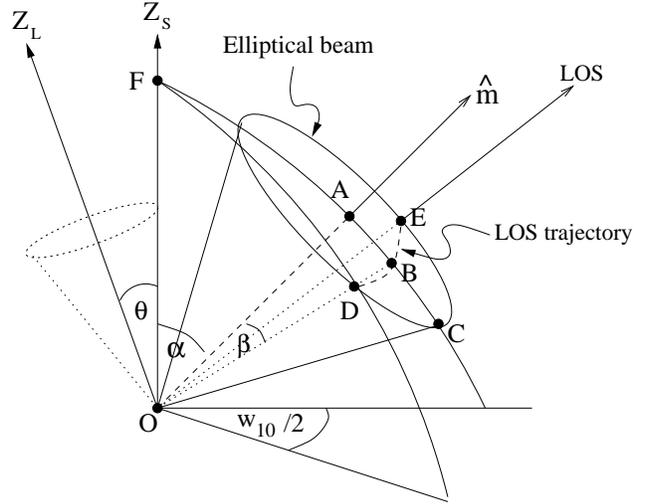}
         \caption{Schematic diagram of the beam geometry of the B pulsar at time $t$. The magnetic, spin, and orbital normal axes are indicated as $\hat m$, $Z_{\rm s}$, and $Z_{\rm L}$, respectively.  The B pulsar's half-opening angle across the semimajor axis of the elliptical beam ($\rho$) is $A \hat O C = 14\fdg3$. In comparison, the {\it effective} half-opening angle ($\rho_{\rm e}$) of the beam at time $t$ is defined to be $A \hat O D$. Due to geodetic precession, the relative motion of the beam with respect to the line of sight (LOS) changes with time, resulting in a variation in the closest approach of the beam to the LOS ($\beta$). Based on our results, $\rho_{\rm e}$ of the B pulsar varies between $5\fdg5$ (non-zero minimum, when the LOS crosses the beam's centre) and $14\fdg3$ (maximum, when the edge of the beam is just grazing the LOS). Following \citet{perera2012}, we fit the pulse widths measured at 10 per cent of the maximum intensity at different MJDs ($W_{\rm 10}$). See \citet{perera2012} for more details about the beam model and pulse width fitting.}
        \label{fig:rhoeff}
       \end{figure}
       \end{center}

The pulse profiles of B have dramatically changed over the five years 
since its discovery. This is because our line of sight cuts through 
different parts of the pulsar emission beam over time due to geodetic spin precession. 
We calculate B's beaming correction factor based on its {\it effective} beam size 
$\rho_{\rm e}$, given a misalignment angle. 
We emphasize that $\rho_{\rm e}$ is different from the pulsar's 
intrinsic beam size ($\rho = 14\fdg3$) 
that represents the angular radius across the semimajor axis of an elliptical beam 
(see fig.\ 5 in Perera et al.\ 2012 \nocite{perera2012} for the schematic plot of B's beam geometry).
The effective beam size is subject to change over time depending on 
how the angle between B's spin axis precesses with respect to our line of sight. 
By definition, $\rho_{\rm e} \le \rho$. 
Fig.\ \ref{fig:rhoeff} illustrates $\rho$ and $\rho_{\rm e}$.

\begin{center}
       \begin{figure} 
         \includegraphics[width=8.5cm]{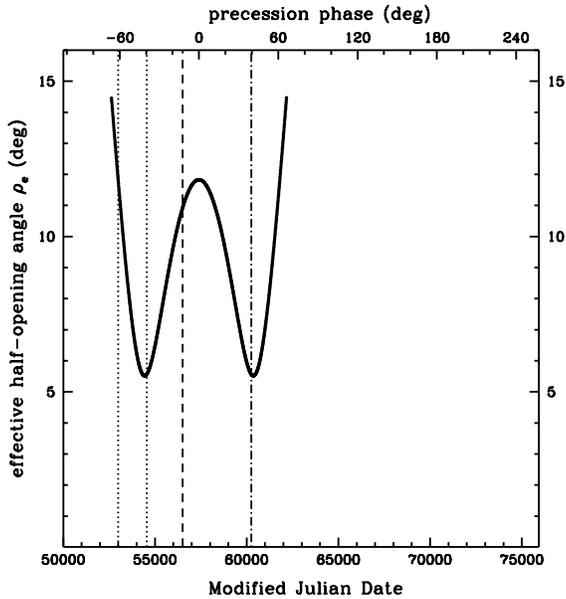}
         \caption{The effective half-opening angle $\rho_{\rm e}$ 
                  of B is shown over its 71-yr geodetic precession period in MJD 
                  (bottom axis) as well as in precession phase 
                  ($\phi_{\rm prec}=61^{\circ}$ at MJD 52997, top axis). 
                  Dotted lines indicate the duration when B was 
                  detectable by the GBT (between MJD 52997 and MJD 54552). 
                  The predictions of B's reappearance time 
                  based on symmetric and single horse-shoe models 
                  \citep{perera2010, perera2012} are shown as 
                  dashed (MJD 56500) and dot--dashed (MJD 60246) lines, respectively.}
        \label{fig:rho}
       \end{figure}
       \end{center}

In order to calculate $\rho_{\rm e}$ of the B pulsar,
we fix the best-fitting values that describe the elliptical beam (including $\alpha$), 
and compute the pulse profile width at 10 percent of the maximum intensity ($W_{10}$) 
by using equations $9-12$ in \citet{perera2012} as a function of time. 
Then we calculate $\rho_{\rm e}$ corresponding to $W_{10}$ by equation 20 in their paper.
Fig.\ \ref{fig:rho} shows the obtained $\rho_{\rm e}$ over the precession 
time-scale of 71 yr, corresponding to $\alpha=61^{\circ}$. 
It varies between $5\fdg5 \le \rho_{\rm e} \le 14\fdg3$.
In early observations, e.g., when our line of sight enters within B's beam,
the apparent beam size of the B pulsar is close to the intrinsic beam size of the full ellipse.
As our line of sight moves upwards to the centre of the beam over time, 
$\rho_{\rm e}$ becomes smaller. We obtain $\rho_{\rm e} = 5\fdg5$ from later observations, 
when our line of sight crosses around the centre of the beam.

We calculate B's effective beaming correction factor\footnote{\citet{ok2010} calculated $f_{\rm b,eff}$ 
by averaging the beaming fraction. If we calculate $f_{\rm b,eff}$ 
for the B pulsar using their equation (10),
fixing $\alpha=61^{\circ}$, we get $f_{\rm b,eff}\sim3.4$.}
($f_{\rm b,eff,B}$) considering the secular change of $\rho_{\rm e}$ and the 95 per cent confidence interval 
for $\alpha$ based on \citet{perera2012}.
We note that the range of $\rho_{\rm e}$ remains the same
between $\alpha = [56^{\circ},77^{\circ}]$ that we consider. 
For a given value of $\alpha$, we randomly select $\rho_{\rm e}$ 
between $[5\fdg5, 14\fdg3]$, assuming a uniform distribution.
We calculate $f_{\rm b,eff}$ by averaging $N=10^{5}$ beaming correction 
factors obtained from equation (\ref{eq:fb}):
\begin{equation}\label{eq:fbeff}
{f_{\rm b,eff}} \equiv \bigl< f_{\rm b,i} (\alpha,\rho_{\rm e,i}) \bigr> = \frac{1}{N}\sum_{i=1}^{N} {f_{\rm b,i}}~.
\end{equation}
Assuming $\alpha=61^{\circ}$, we obtain the reference beaming correction 
factor for the B pulsar to be $f_{\rm b,eff}=3.7$.

The beam size of canonical pulsars with spin periods $P_{\rm s}> 0.1$ s
can be estimated from its spin period by 
the empirical relation, i.e., $\rho(\pspin) \propto \pspin^{-0.5}$ (e.g.\ Kramer et al.\ 1998; Tauris \& Manchester 1998 and refences therein).\nocite{tm98,krameretal98}
This is based on a circular beam model where 
the half-opening angle of the beam $\rho$ is assumed to be constant over time. 
This relation is useful to estimate the beam size of pulsars 
with simple and stable pulse profiles, e.g., typical canonical pulsars, 
where our line of sight always cuts through the same part of the beam.
However, the $\rho$--$\pspin$ relation can fail 
to describe the beam function of pulsars like B, 
when the pulse profile (i.e., the beam size) is time dependent.

In Fig.\ \ref{fig:fbeff}, we compare the estimated $f_{\rm b,eff}$ based on the elliptical 
beam model considering the plausible range of $\rho$ and $\alpha$ (solid), 
with $f_{\rm b}$ obtained by a fixed $\rho=3\fdg2$ obtained from the $\rho$--$\pspin$ relation (dashed)
between $\alpha=(0^{\circ},90^{\circ}]$. 
The effective beaming correction factor is robust within 
 the 95 per cent interval of $\alpha$ between $[56^{\circ},77^{\circ}]$.
The beaming correction factor based on the $\rho$--$\pspin$ relation 
is overestimated, regardless of the value of $\alpha$, from what is preferred by the more realistic elliptical beam model.

As for the reference beaming correction factor for A, 
we follow similar steps described in \citet{ferdman2013}.  
However, we use pulse profiles at a more conservative 
5 per cent intensity level instead of the
25 per cent used by \citet{ferdman2013} as they allow greater sensitivity
to subtle changes in the pulse profiles. By fitting each beam of the two-pole model independently to the observed pulse profiles, we obtain  $\alpha=88\fdg2$, $\rho_{\rm 1}=27\fdg2$ and $\rho_{\rm 2}=32^{\circ}$. This implies $f_{\rm b,eff,A}\simeq2$ and we use this as a reference value for A in this work. The details of the pulse profile analysis for the A pulsar will be presented in a separate paper (Perera et al.\ 2014)\nocite{pereraprep}.

\subsection{Bright and Weak Orbital Phases}\label{sec:zeta}

The orbital longitudes of the bright and weak phases given in table 1 in \citet{perera2010} 
imply that each phase is observable for only $\sim 10-15$ per cent of B's full orbital phase. 
The orbital longitudes of {BP1, BP2, WP1, and WP2} are {$190^{\circ} - 235^{\circ}$, 
$260^{\circ} - 300^{\circ}$, $340^{\circ} - 30^{\circ}$, and $80^{\circ} - 130^{\circ}$, 
respectively. This is consistent with the earlier observations made by the Parkes 
telescope at 1390 MHz \citep{burgay2005}. 

We introduce a dimensionless factor $\zeta$ that represents 
the fractional time when the B pulsar is detectable 
in orbit, in other words, the total area of B's orbit divided by the area of 
elliptical sectors of $BP1+BP2+WP1+WP2$. Considering the orbital longitudes of all detectable 
phases, B was observable over only about half of the orbit and $\zeta_{\rm B} \sim 1.9$. 
The GBT observations imply that the combined fraction of bright phases decreased over time. 
As we use the observable orbital longitudes measured during early observations when 
B appears brighter than later in time, $\zeta_{\rm B}\sim1.9$ is conservative.

The A pulsar is detectable over all phases of the 2.45 h orbit except for the 30 s eclipse 
\citep[e.g.,][]{J0737Adiscovery}. 
Therefore, we can safely assume $\zeta_{\rm A}=1$. All pulsars found in 
the known NS$-$NS binaries, such as PSR B1913+16, have $\zeta=1$, except the B pulsar.

B's beaming direction keeps changing mainly due to the geodetic precession, but the time-evolution of a pulsar can be fully understood by (a) its beam shape and orientation as well as (b) all effects which affect the direction of the beam. In the rate equation, we treat $\tau_{\rm life}$ as a constant parameter for a selected binary. Then we correct $N_{\rm psr}$ by multiplying the averaged beaming correction factor $f_{\rm b,eff}$ (obtained in \S 3.3) and the $\zeta$ parameter. This treatment can be justified by the most recent interpretation for the orbital flux variation of the B pulsar, namely, that B's radio emission is always bright but that its radio beam is deflected into our line of sight by A's wind during only two `bright phases' in its orbit. In addition, we assume that there are the same numbers of B-like pulsars pointing towards and away from the Earth. We can therefore use equations (1) and (4) as they are, only replacing $N_{\rm psr} f_{\rm b,eff}$ by $N_{\rm psr} f_{\rm b,eff} \zeta$ for the B pulsar.”

\subsection{Effective Lifetime}\label{sec:lifetime} 

An effective lifetime of an NS$-$NS binary, \tlife, is defined 
\begin{eqnarray} \label{eq:lifetime}
\tau_{\rm life} &\equiv& \tau_{\rm age}+\tau_{\rm obs} \\
&\equiv& \text{min}(\tau_{\rm c},\tau_{\rm c}[1-(P_{\rm birth}/P_{\rm s})]^{n-1}) + \text{min}(\tau_{\rm mrg},\tau_{\rm d})~ \nonumber,
\end{eqnarray}
where $\tau_{\rm age}$ is the current age of the pulsar,
determined by its current spin period and 
period derivative (with an assumption on its surface magnetic field), 
and $\tau_{\rm obs}$ represents the binary's remaining {\it observable} time-scale 
from the current epoch. 

The characteristic age \tc$\equiv \pspin/(n-1){\dot P}_{\rm s}$ is 
typically considered as $\tau_{\rm age}$ 
for non-recycled pulsars with spin periods of $\sim 1$ s like B,
where $n$ is a magnetic braking index.
For recycled pulsars such as the A pulsar, however,
we calculate their effective spin-down ages by \tc$[1- (P_{\rm birth}/P_{\rm s})]^{n-1}$. 
This is based on an assumption that  
current spin periods of recycled pulsars are comparable to 
their birth periods $P_{\rm birth}$s \citep{spindown99}. 
We consider the effective spin-down age as the reference age estimate 
for all recycled pulsars used in this work.
For non-recycled pulsars, we choose their characteristic age. 
For simplicity, we assume that all pulsars in merging binaries 
have surface dipole magnetic fields with magnetic braking index $n=3$ 
and no magnetic field decay.

The Double Pulsar provides us with two age 
constraints from the A and B pulsars. 
The characteristic age of the B pulsar is $\sim 50$ Myr.
However, \citet{J0737age} suggested that the age 
of the B pulsar is likely to be between 50 and 190 Myr, where 
the upper limit is favoured by a model involving interactions 
between A's wind and B's magnetosphere (model 4). 
We use A's effective spin-down age ($\sim 140$ Myr) as the current age of the Double Pulsar, assuming independent spin-down history for A and B for simplicity.

The remaining lifetime of the binary $\tau_{\rm obs}$ 
used in the empirical method concerns the detectability of pulsar(s) 
in the binary by radio pulsar surveys. It is determined by 
\td, the radio emission time-scale or 
the so-called death-time\footnote{As pointed out in \citet{ok2010}, 
there is $\sim 70$ per cent uncertainty in $\tau_{\rm d}$ of the B pulsar. 
If the gap potential for B is $V_{\rm g} \sim 10^{12}$ V 
\citep{spindown2006}, the radio emission time-scale of B can 
be as long as $\sim 90$ Myr. The uncertainty in the peak rate estimate attributed 
to B's radio emission time-scale is $\sim 15$ per cent.} 
(e.g., Chen \& Ruderman 1993)\nocite{cr93}, 
or \tmrg~that is a merging time-scale of the binary due to GW emission \citep{pm63}.
For the Double Pulsar, $\tau_{\rm obs}$ is determined by
the B's radio emission time-scale of 40 Myr.
Based on what is described above, 
the effective lifetime of the Double Pulsar is estimated to be 
$\tau_{\rm age, A} + \tau_{\rm d,B}=180$ Myr. 

For comparison, we note that \citet{kalogera2004} used 185 Myr as the lifetime of the Double Pulsar,
which is the sum of the effective spin-down age estimate for the A pulsar ($\sim100$ Myr) 
and the binary merger time-scale ($\sim 85$ Myr).
Their age estimate for A is based on ${\dot P}_{\rm s}=2.3\times10^{-18}$ ss$^{-1}$ 
measured by \cite{J0737Adiscovery} when A was discovered.
\citet{ok2010} and this work adopt ${\dot P}_{\rm s}=1.74\times10^{-18}$ ss$^{-1}$ from the follow-up timing observations \citep{kramer2006}.

We assume that the epochs of observation as well as the beam 
directions of any B-like pulsars are random. 
This implies that there are equal numbers of pulsars 
beaming towards our line of sight at any epoch, 
and hence, \tlife~of B or the Double Pulsar is not 
affected by its geodetic precession time-scale of 71 yr. 
Applying the same equivalent assumption to the PSR B1913+16-like 
pulsar population, their lifetime is defined to be 
$\tau_{\rm age} + \tau_{\rm mrg}=370$ Myr, even though
PSR B1913+16 is expected to move away from our 
line of sight around 2025 and will return in 2220 
\citep[e.g., ][]{kramer1998B1913,kramerprecession2010}.

       \begin{center}
         \begin{figure} 
           \includegraphics[width=8.5cm]{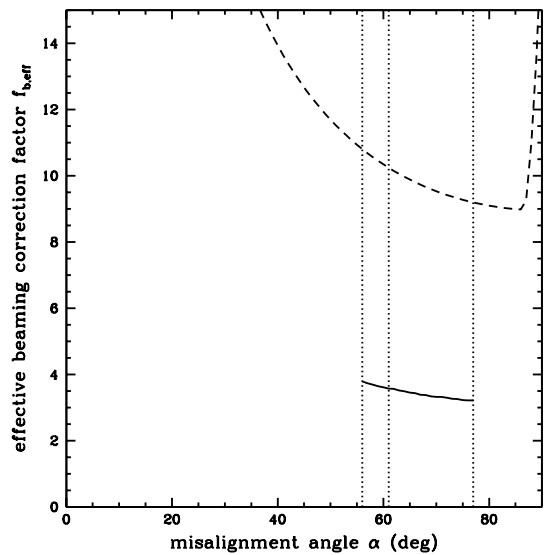} 
           \caption{We compare B's effective beaming correction 
                    factor $f_{\rm b,eff}$ (solid) 
                    and $f_{\rm b}$ (dashed) based on the empirical 
                    $\rho \propto P_{\rm s}^{-0.5}$ relation. The dotted vertical 
                    lines are the suggested magnetic misalignment angles within the 95 per cent confidence interval 
                    error given by \citet{perera2012}, i.e., $\alpha=56^{\circ}$ (left), 
                    $61^{\circ}$ (centre), and $77^{\circ}$ (right). 
                    Our reference value is $f_{\rm b,eff}=3.7$, 
                    obtained with the best-fitting $\alpha=61^{\circ}$ of the 
                    elliptical beam model.}
           \label{fig:fbeff}
        \end{figure}
       \end{center}

\section{The Galactic NS$-$NS Merger Rate Estimates} \label{sec:rates}

In this section, we derive \prate~for the Double Pulsar 
using both A and B and calculate \grate~considering 
PSRs B1913+16, B1534+12, and the Double Pulsar. 

Table \ref{tab:allparameters} summarizes reference parameters 
used for each NS$-$NS binary.
We note that all the beaming correction factors are constrained by pulsar observations.
For PSRs B1913+16 and B1534+12, 
we adopt $\rho$ and $\alpha$ estimated by polarization measurements 
(see Kalogera et al.\ 2001\nocite{knst} for further details). 
For $N_{\rm pop}$ and $C$, we show
rounded values to the nearest hundreds and thousands digits.
However, we show $N_{\rm psr}$ for all pulsars as obtained 
from the Monte Carlo simulations including the PALFA surveys.     
In addition to reference values for A and B (indicated as REF),
we also show parameters and results for a case with $f_{\rm b,eff,A}=6$ for comparison.

\citet{kkl2003} showed that the likelihood of detecting a pulsar 
like one of the known pulsars follows the Poisson distribution. 
In this section, we focus on the A and B pulsar-like populations (i.e., $i=$ A,B), but
this likelihood can be applied to any pulsar binaries found in the Galactic disc, 
containing one detectable pulsar:
\begin{equation}\label{eq:likelihood1}
      {\cal L}_{i}(D_{i} | {\lambda}_{i} X) = \frac{{{\lambda}_{i}}^{D_{i}} {\rm e}^{-{{\lambda}_{i}}}} {D_{i} !}~,
\end{equation}
where $D_{i}$ is the number of the observed sample like the pulsar $i$
(data), ${\lambda}_i$ is the mean of the Poisson distribution (our hypothesis), 
and $X$ is the model assumption. 
Applying this likelihood to Bayes' theorem (i.e.\ ${\rm posterior} \propto {\rm likelihood} \times {\rm prior}$), the posterior PDF for the pulsar population $i$ is obtained to be
$P_{i}(\lambda_{i} | D_{i} X) \equiv P_{i}(\lambda_{i}) = \lambda_{i} {\rm e}^{-{\lambda_{i}}}$, where $D_{i}=1$. As pointed out by \citet{kkl2003}, 
when $D=1$, the maximum ($P(\lambda)={\rm e}^{-1}$) occurs at $\lambda=1$.
For simplicity, we omit the conditions of the PDFs hereafter. 
All posteriors are conditional PDFs, given $D_{i}=1$, 
and are based on our reference model.
Equation (\ref{eq:rate}) is based on the likelihood\footnote{\citet{kkl2003} 
used $N_{\rm obs}$ and $N_{\rm tot}$ instead of $D$ and $N_{\rm psr}$. 
Equation (\ref{eq:likelihood1}) is the same with equation 7in their paper.} 
(see section 5.1 in their paper for derivation).

Our modelling for A and B is implicitly based on the fact that A and B pulsars 
belong to the Double Pulsar. The flux degradation factor for A (due to its orbital motion) 
and the pulse width of B (changed due to geodetic precession) depend on the masses 
of two neutron stars as well as binary properties.
Therefore, the likelihood of detecting a binary similar to the Double Pulsar can be written 
as a product of the likelihoods of detecting A and B, which are the same with equation (5):
\begin{eqnarray}\label{eq:likelihood2}
{{\rm likelihood}_{\rm J0737}}  & \equiv & {\rm likelihood_A}~ \times~ {\rm likelihood_B}~, \nonumber \\
                                & = & {{\lambda}_{\rm A}} {\rm e}^{-{{\lambda}_{\rm A}}} \times {{\lambda}_{\rm B}}{\rm e}^{-{{\lambda}_{\rm B}}}~.
\end{eqnarray}
The posterior of detecting a binary like the Double Pulsar (consisting of A and B pulsars) is therefore 
\begin{eqnarray}\label{eq:posterior}
P(\lambda_{\rm J0737}| D_{\rm J0737} X) & \equiv & P(\lambda_{\rm A},\lambda_{\rm B} | D_{\rm A} D_{\rm B} X) ~\nonumber \\ 
& = & {{\lambda}_{\rm A}}  {{\lambda}_{\rm B}} {\rm e}^{-({{\lambda}_{\rm A}+{\lambda}_{\rm B}})}~, 
\end{eqnarray}
where $D_{\rm A}=D_{\rm B}=1$, and therefore, $D_{\rm J0737}=1$ in this work. 
We note that it is impossible 
to directly calculate $P(\lambda_{\rm J0737}| D_{\rm J0737} X)$, as the detection of 
the Double Pulsar (i.e., counting of $D_{\rm J0737}$) obtained only when 
both A and B are detected by pulsar observations, independently. 
Therefore, what we can calculate from the pulsar observations 
is $P(\lambda_{\rm A},\lambda_{\rm B} | D_{\rm A} D_{\rm B} X)$.

\begin{table} 
\caption{Reference parameters and results of the NS$-$NS binaries considered in this work. 
The correction factor taking into account detectable orbital phase $\zeta$ 
is assumed to be unity, except the B pulsar ($\zeta=1.9$).
The results corresponding to $f_{\rm b,eff}=6$ for the A pulsar are listed for comparison.
See the text for the definition of all parameters.}
\label{tab:allparameters}
\footnotesize
\begin{center}
\begin{tabular*}{8.5cm}{llllllc} 
\hline
PSR name & $f_{\rm b,eff}$ & $\delta$ & $N_{\rm psr}$ & $N_{\rm pop}$ & \tlife& $C$ \\
         &                 &          &               &               & (Gyr) & (kyr) \\
\hline
 A (REF)   & 2    & 0.27  & 907 & 1800  & 0.18  &  100 \\
 A         & 6    & 0.27  & 907 & 5400  & 0.18  &   30 \\
 B (REF)   & 3.7  & 0.013 & 213 & 1500  & 0.18  &  120 \\
 B1913+16  & 5.72 & 0.169 & 392 & 2200  & 0.37  &  170 \\
 B1534+12  & 6.04 & 0.04  & 253 & 1500  & 2.93  & 1900 \\
\hline
\end{tabular*}
\end{center}
\end{table}

Due to the different observational biases, 
$N_{\rm psr,A}$ and $N_{\rm psr,B}$ are not necessarily the same. 
Based on our results, the A pulsar is more likely to be detected 
than the B pulsar ($N_{\rm psr,B} < N_{\rm psr,A}$).
If we correct the observational biases perfectly, however,
the total number of the Double Pulsar ($N_{\rm pop,J0737}$) in the Galactic disc 
estimated by A and that based on B are to be the same:
\begin{equation}\label{eq:npopAB}
N_{\rm pop,A} = N_{\rm pop,B} \equiv N_{\rm pop,J0737}~.
\end{equation}
As shown in Table \ref{tab:allparameters},
the population sizes of the Double Pulsar estimated by A ($N_{\rm pop,A}=1400$) 
and B ($N_{\rm pop,B}=1500$), respectively, from reference parameters are consistent.

Recalling $s=1/N_{\rm psr}$ from the linear relation $N_{\rm det}=sN_{\rm psr}$ 
and using equation (\ref{eq:npopAB}),
we can express $\lambda_i$ ($i=$ A,B) 
as a function of $N_{\rm pop,J0737}$ and the correction factors we discussed earlier.
\begin{equation}\label{eq:lambda}
\lambda_i   =    \frac{s_i N_{\rm pop,J0737}}{f_{\rm b,eff,i} \zeta_i} = \frac{N_{\rm pop,J0737}}{f_{\rm b,eff,i} \zeta_i N_{\rm psr,i}} \equiv \frac{N_{\rm pop,J0737}}{c_i}~,
\end{equation}
where the constant $c_i$ is introduced for simplicity. 
Note $c_i=C_i/\tau_{\rm life,i}$ ($i=$ A,B) and the uncertainty in $N_{\rm pop}$ is attributed to 
the pulse profile change (of B) and the details of beam functions (of both A and B).
The PDF for the population size of the Double Pulsar
$P(N_{\rm pop,J0737})$ can then be obtained by changing of variables from equation (\ref{eq:posterior}):
\begin{equation}\label{eq:npoppdf}
P(N_{\rm pop,J0737})={\frac{(c_{\rm A}+c_{\rm B})^{3}}{2}} {{N_{\rm pop}}^{2}} {\rm e}^{-(c_{\rm A}+c_{\rm B}) N_{\rm pop,J0737}}~.
\end{equation}
It is straightforward to calculate ${\cal P}({\cal R}_{\rm J0737})$ applying a chain rule.
\begin{eqnarray}\label{eq:ABrate}
{\cal P}({\cal R}_{\rm J0737}) &=& {\cal P}(N_{\rm pop, J0737}) \left | \frac{{\rm d}N_{\rm pop, J0737}}{{\rm d}{\cal R}_{\rm J0737}} \right | \nonumber \\
                               &=& \frac{(C_{\rm A}+C_{\rm B})^{\rm 3}}{2}{{\cal R}^{2}}_{\rm J0737} {\rm e}^{-(C_{\rm A}+C_{\rm B}){\cal R_{\rm J0737}}} \nonumber \\
                               & \equiv & {\cal P}_{\rm 1}({{\cal R}_{\rm 1}})~.
\end{eqnarray}

We emphasize that equations (\ref{eq:likelihood2})$-$(\ref{eq:ABrate}) 
can be used only when both NSs in the binary are detected as radio pulsars 
and their observational biases are reasonably well understood (i.e., the rate coefficients of both pulsars 
should be well constrained and comparable). 
When there is only one detectable pulsar in the binary available for the rate calculation, 
one can follow the steps described in \citet{kkl2003} that result in equation (\ref{eq:rate}). 
Even though the B pulsar has been known since 2004, due to the lack of information 
to model this pulsar, previous works used only the A pulsar's 
properties that are better understood.

If all selection effects are properly accounted for, the joint PDF ${\cal P}({\cal R}_{\rm J0737})$ (based on both A and B) should have the same peak rate estimate (${\cal R}_{\rm peak}$) predicted by the original rate equation based on the single pulsar (either A or B). In other words, the peak rate estimates of equations (\ref{eq:rate}) and (\ref{eq:ABrate}) occur at ${\cal R}_{\rm peak}=1/C$, i.e., ${\rm d}{\cal P}({\cal R})/{\rm d}{\cal R}=0$ at ${\cal R}_{\rm peak}=1/C$ where $C=C_{\rm A}=C_{\rm B}$. 
The equality in rate coefficients is satisfied when selection effects for A and B pulsars are correctly applied. As shown in Table 2, our results reasonably satisfy this condition.
Based on the consistency in model assumptions and derived rate equations, 
our results can be directly compared with previous works based on only the A pulsar \citep[e.g.,][]{kalogera2004}.

In Fig.\ \ref{fig:npop}, we plot 
individual PDFs for $N_{\rm pop}$ based on A ($P(N_{\rm pop, A})$, dotted) 
and B ($P(N_{\rm pop, B})$, dashed), overlaid with 
$P(N_{\rm pop, J0737})$ (solid). 
For our reference model, $P(N_{\rm pop, A})$ and $P(N_{\rm pop, B})$ are consistent.
Note that $P(N_{\rm pop, J0737})$ has narrower width than those of individual PDFs, as expected.
Based on the combined $P(N_{\rm pop, J0737})$, 
we expect there are $\sim 1500^{+4000}_{-1000}$ 
systems like the Double Pulsar in the Galactic disc at 95 per cent confidence. 
If we assume $f_{\rm b,eff,A}=6$, 
$N_{\rm pop, J0737} \sim 2000^{+5000}_{-1900}$. 

Finally, we calculate the PDF of Galactic NS$-$NS merger rate estimates 
${\cal P}_{\rm g}({\cal R}_{\rm g})$.
In order to do this, we need a combined PDF 
based on PSRs B1913+16 and B1534+12, ${\cal P_{\rm 2}}({\cal R_{\rm 2}})$, 
which is derived by \citet{kkl2003} as follows
\begin{eqnarray} \label{eq:p2}
  {\cal P}_{\rm 2}({\cal R}_{\rm 2}) &=&  {\Bigl( \frac{{C_{\rm 1913}}{C_{\rm 1534}}}{{C_{\rm 1534}}-{C_{\rm 1913}}}\Bigr)^{2}} \Bigl[ {{\cal R}_{\rm 2}} {\bigl({ {\rm e}^{-C_{\rm 1913}{\cal R}_{\rm 2}} + {\rm e}^{-C_{\rm 1534}{\cal R}_{\rm 2}} }\bigr)} \nonumber \\
  & - & \Bigl( {\frac{2}{{C_{\rm 1534}}-{C_{\rm 1913}}} \Bigr)} {\bigl({ {\rm e}^{-C_{\rm 1913}{\cal R}_{\rm 2}} - {\rm e}^{-C_{\rm 1534}{\cal R}_{\rm 2}} }\bigr)} \Bigr]~.
\end{eqnarray}
As described in section 5.2 in \citet{kkl2003}, 
we can calculate the PDF of Galactic NS$-$NS merger rate estimates 
from equations (\ref{eq:ABrate}) and (\ref{eq:p2}):
\begin{eqnarray}\label{eq:tot}
{\cal P}_{\rm g}({\cal R}_{\rm g}) 
& = & \int_{{\cal R_{\rm -}}={-{\cal R_{\rm g}}}}^{{\cal R}_{\rm -}=+{\cal R_{\rm g}}} d{\cal R_{\rm -}} \frac{1}{2}{\cal P}_{\rm 1}({\cal R_{\rm 1}}) {\cal P}_{\rm 2}({\cal R_{\rm 2}}) \nonumber \\
& =  & {\frac{C_{\rm 1913}^{\rm 2}C_{\rm 1534}^{\rm 2} (C_{\rm A} + C_{\rm B})^{\rm 3}}{4}} \int_{{\cal R}_{\rm -}} d{\cal R_{\rm -}} {{\cal R}_{\rm -}^{2} {\rm e}^{-(C_{\rm A}+C_{\rm B}) {\cal R_{\rm -}}}} \nonumber \\
&    & \Bigl[ {\cal R_{\rm g}} \left({\rm e}^{-C_{\rm 1913} {\cal R_{\rm g}}} + {\rm e}^{-C_{\rm 1534} {\cal R_{\rm g}}}\right)  \nonumber \\
& - & {\frac{2}{C_{\rm 1534}-C_{\rm 1913}} } \left({\rm e}^{-C_{\rm 1913} {\cal R_{\rm g}}} - {\rm e}^{-C_{\rm 1534} {\cal R_{\rm g}}} \right) \Bigr] ~,
\end{eqnarray} 
where ${\cal R_{\rm -}} \equiv {\cal R}_{\rm 1} - {\cal R}_{\rm 2}$, 
and ${\cal R_{\rm g}} \equiv {\cal R}_{\rm 1} + {\cal R}_{\rm 2}$.
Based on the results given in Table 2, $C_{1913} < C_{\rm A} + C_{\rm B} < C_{1534}$. 

Based on our merger rate estimates,
we calculate the GW detection rate for NS$-$NS inspirals with
ground-based interferometers by 
\begin{equation}\label{eq:rdet}
{\cal R}_{\rm det}={\cal R}_{\rm g} \times N_{\rm G}~,
\end{equation} 
where $N_{\rm G}\equiv (4{\rm \pi}/3) (d_{\rm h,Mpc}/2.26)^{3} (0.0116)$ 
is the number of Milky Way equivalent galaxies that would contain 
NS$-$NS binaries within the detection volume of 
the advanced ground-based GW detectors 
and $d_{\rm h,Mpc}=445$ Mpc is the horizon distance 
for NS$-$NS inspirals with the advanced LIGO-Virgo network \citep{ligoratespaper2010}. 
See equation (5) and table 5 in their paper for more details.

Fig.\ \ref{fig:ratesall} shows 
${\cal P}_{\rm g}$(${\cal R}_{\rm g}$) (solid) along with 
the individual PDFs of rate estimates for PSRs B1913+16 (dotted) 
and the Double Pulsar (short dashed). 
Although we consider PSR B1534+12 in the rate calculation, 
we do not show the PDF for PSR B1534+12 in 
Fig.\ \ref{fig:ratesall} for clarity.
Throughout this paper, we use equation (\ref{eq:rate}) to calculate \prate~for 
PSRs B1913+16 and B1534+12 as there is only one known pulsar component in these binaries.
The PDF for the Double Pulsar is obtained from equation (\ref{eq:ABrate}) constrained by both $C_{\rm A}$ and $C_{\rm B}$.

We note that, although we assume $N_{\rm pop,A}=N_{\rm pop,B}=N_{\rm pop,J0737}$ to calculate
equations (\ref{eq:npoppdf}) and (\ref{eq:ABrate}),
we incorporate individually estimated $C_{\rm pop,A}=100$ kyr and $C_{\rm pop,B}=120$ kyr to plot 
Figs \ref{fig:npop} and \ref{fig:ratesall} (see Table \ref{tab:allparameters}). 
The lifetime of the Double Pulsar is estimated to be 180 Myr as described in \S \ref{sec:lifetime}.

Our reference model implies ${\cal R_{\rm g}}=21_{\rm -14}^{+28}$ Myr$^{-1}$ 
and ${\cal R}_{\rm det}=8_{\rm -5}^{+10}$ yr$^{-1}$. If we assume $f_{\rm b,eff,A}=6$, 
as used in some of the previous work, we obtain ${\cal R_{\rm g}}=26_{\rm -17}^{+33}$ 
Myr$^{-1}$, and ${\cal R}_{\rm det}=10^{+12}_{-6}$ yr$^{-1}$. 
All results in this section are given at the 95 per cent confidence interval.

       \begin{center}
       \begin{figure}
         \includegraphics[width=8.5cm]{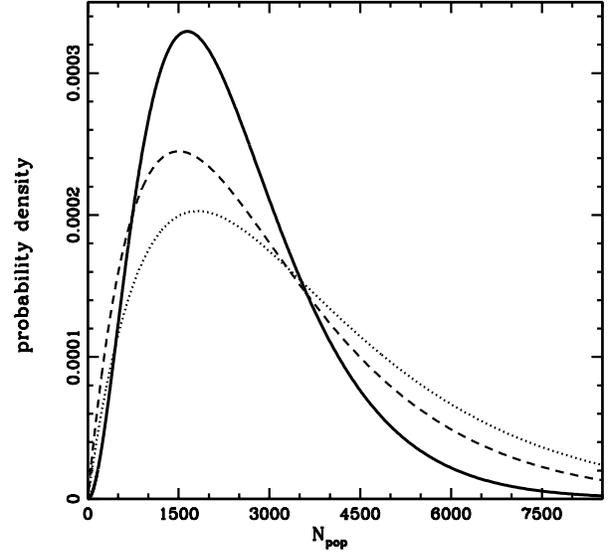}
         \caption{${\cal P}$($N_{\rm pop}$) for the Double Pulsar (solid) 
                  and the individual PDFs for A (dotted) and B (dashed) 
                  based on our reference model. 
                  At 95 per cent confidence, we expect there are 
                  $1500^{+4000}_{-1000}$ NS$-$NS binaries similar to 
                  the Double Pulsar in the Galactic disc.}
         \label{fig:npop}
       \end{figure}
       \end{center}

\section{Discussion}\label{sec:discussion}

In this work, we consider four pulsars 
(PSRs 1913+16, 1534+12, J0737$-$3039A and J0737$-$3039B) that
represent three NS$-$NS binaries in the Galactic disc,
following similar steps described by Kim et al.\ \nocite{kkl2003}(2003, see sections $2-4$ in their paper for details).
For the first time, we calculate the merger rate of the Double Pulsar using
the non-recycled B pulsar based on the 5-yr GBT observations.
This allows us to derive ${\cal P}({\cal R}_{\rm J0737})$ 
for the Double Pulsar based on both A and B pulsars (equation \ref{eq:ABrate}). 
Assuming the three pulsar binaries fully represent the Galactic NS$-$NS population,
we calculate \grate~as well as the corresponding GW detection rates 
for advanced ground-based GW detectors. 

Based on our reference model, 
the Galactic NS$-$NS merger rate is 
${\cal R}_{\rm g}=21_{\rm -14}^{+28}$ Myr$^{-1}$ and  
$21_{\rm -17}^{+40}$ Myr$^{-1}$ 
at 95 and 99 per cent confidence intervals, respectively.
The peak rate estimate is 
smaller than what previously known (e.g., Kalogera et al.\ 2004\nocite{kalogera2004}).
This is mainly due to the smaller beaming correction factors 
estimated for A and B. 
In addition, the single discovery of an NS$-$NS binary 
from the PALFA precursor survey that has a large 
field of view and better sensitivity than previous 
surveys is attributed to the estimated \npsr~of each pulsar 
being smaller by a factor $1.5-1.7$ from those given in 
\citet{ok2010}.
We note that the contributions from the Double Pulsar and 
PSR B1913+16 are comparable
and no single binary dominates the Galactic NS$-$NS merger rate.

Motivated by the independent constraints from the B pulsar 
such as $P(N_{\rm pop})$, we believe that A's beam is likely to be wider 
than those of PSRs B1913+16 ($\rho=12\fdg4$) and B1534+12 ($\rho=4\fdg87$).
Furthermore, the long-term observations of PSRs B1913+16, J1141$-$6545 
and the Double Pulsar (through A and B) imply that 
individual pulsar beam patterns can be quite different. 
In this work, we consider the three 
pulsar binaries with the best observational constraints.

Systematic uncertainties related to the pulsar population modelling 
(e.g., distribution of pulsars in the Galactic disc, radio pulsar luminosity distribution,
current age of the Double Pulsar) are studied by
 \nocite{kkl2003,kkl2010}Kim et al.\ (2003, 2010), and \citet{ok2010}. 
In this work, we examine systematic uncertainties in the rate estimates,
focusing on the two relatively least constrained parameters for the Double Pulsar,
$\tau_{\rm life}$ and $N_{\rm psr,B}$ within the plausible range.
The lifetime for the Double Pulsar ranges between
$\tau_{\rm life} \sim 90$-$230$ Myr, and $N_{\rm psr,B} \sim 190$-$270$ 
(or $N_{\rm pop,B} \sim ~ 1300$-$1900$) 
attributed to uncertainties in B's radio emission time-scale (\S \ref{sec:lifetime}) 
and different duty cycles due to geodetic precession (\S \ref{sec:weq}). 
We consider two cases, assuming parameters at the extremes allowed by observation:
(a) B-like pulsars with broad pulse profile ($\delta=0.03$, $N_{\rm pop} \sim 1900$) 
and longest plausible lifetime of $\tau_{\rm life,J0737}=230$ Myr, 
and (b) those with narrow pulse profile ($\delta=0.005$, $N_{\rm pop} \sim 1300$)
and the reference binary lifetime $\tau_{\rm life,J0737}=180$ Myr.
All other parameters are fixed to our reference model.
For the parameters we explore,
the peak values of ${\cal R}_{\rm g}$ range between $\sim 17$ and 27 Myr$^{-1}$.
The lower and upper limits at 95 per cent confidence are obtained to be 
${\cal R}_{\rm g} \sim 5$ and $\sim 60$ Myr$^{-1}$, respectively.
Although it is not very likely, 
if the lifetime of the Double Pulsar is as short as 90 Myr 
motivated by B's characteristic age, 
${\cal R}_{\rm g} = 36_{\rm -26}^{+59}$ Myr$^{-1}$ at 95 per cent confidence.

The B pulsar was detected by a follow-up observation of the A pulsar.
As described in \S \ref{sec:model}, we calculate $N_{\rm psr}$ by survey simulation, 
where pulsar detection is determined by comparing a model pulsar's radio flux density 
and a survey's sensitivity using the pulsar radiometer equation \citep{radiometer}.
Searching for a companion of a known pulsar in a binary effectively
increases the number of telescope pointings to the location of the binary, but this does not alter the radiometer equation itself.
By assuming the same survey integration times, we underestimate the integration time, 
and hence the sensitivity\footnote{The survey sensitivity is proportional to (integration time)$^{-1/2}$.}, 
of the Parkes High-Latitude pulsar survey for B \citep{J0737Bdiscovery,phsurvey}. 
This implies that $N_{\rm psr,B}$ for this survey is overestimated. Other surveys are not affected. 
It is however difficult to quantitatively assess the uncertainty in $N_{\rm psr,B}$ attributed to 
the treatment of follow-up observations in this work. 
As a rough estimate, we compare our reference result and the minimum expected 
$N_{\rm psr,B}$ by applying the condition $1 < f_{\rm eff,A}$. 
The condition implies that A's beam size is less than $4{\rm \pi}$. 
We rewrite the condition using equation (8) as follows: 
$N_{\rm psr,A}/\zeta_{\rm B}/f_{\rm eff,B} < N_{\rm psr,B}$.
Plugging numbers from Table 2 into this relation, we find that our reference value for $N_{\rm psr,B}$ 
is overestimated less than a factor 2 ($129 < N_{\rm psr,B}$).

Based on recent mass measurements, it is likely that the companion of PSR J1906+0746 is another NS \citep{vanleeuwen2015}. We do not include PSR J1906+0746 in the rate calculation, however, because its beam function is not constrained.
Assuming this pulsar is another NS$-$NS binary, 
we discuss its possible contribution to the Galactic NS$-$NS merger rate using properties of the detected non-recycled pulsar.
If we take $N_{\rm psr} \sim 200$ and $f_{\rm b,eff,J1906} \sim 3-5$ (based on the empirical 
$\rho$--$\pspin$ relation) given by \citet{ok2010}, 
we obtain $N_{\rm pop} \sim 600 - 1000$ for the PSR J1906+0746-like pulsar population. 
As PSR J1906+0746 is detectable at all orbital phases \citep{kasianthesis}, 
we can assume $\zeta=1$. The rate coefficient of this pulsar is $\sim 80 - 130$ kyr 
based on the estimated $N_{\rm pop}$ and its lifetime of $\sim 80$ Myr.
The lifetime of PSR J1906+0746 is determined by its characteristic age and radio emission time-scale 
with no magnetic field decay. As of 2008, PSR J1906+0746 shows only mild pulse 
profile changes compared with those of B \citep{desvignes2008J1906}. 
Our assumptions imply that the rate coefficient of PSR J1906+0746 could be the smallest among the
known NS$-$NS binaries ($C_{\rm 1906} < C_{\rm 1913}$) depending on the beaming correction factor. 
In this case, the contribution of PSR J1906+0746 to the Galactic NS$-$NS merger rate is expected to be significant. 
The Galactic NS$-$NS merger rate including PSR J1906+0746\footnote{Due to the uncertainty in the beam function of PSR J1906+0746, 
its $f_{\rm b,1906}$ is not constrained. Therefore, we provide only the expected merger rate including the pulsar in this work.},
assuming $f_{\rm b,1906}=3.4$ \citep{ok2010}, is expected to be ${\cal R}_{\rm g}\sim40$ Myr$^{-1}$.

The main uncertainty with PSR J1906+0746 is attributed to its beaming correction factor.
\citet{ok2010} estimated the pulsar's beaming correction factor, assuming it follows the $\rho \propto P_{\rm s}^{-0.5}$ relation. 
It seems mildly recycled pulsars, those which often found in NS$-$NS binaries, show deviations from the simple $\rho \propto P_{\rm s}^{-0.5}$ relation.
For example, the beaming correction factors for PSRs B1913+16 and B1534+12 (2.26 and 1.89, respectively) 
calculated by the $\rho \propto P_{\rm s}^{-0.5}$ relation are smaller by factors of 2$-$3
comparing to the measurements (5.72 and 6.04, respectively; see table 1 in O'Shaughnessy \& Kim 2010\nocite{ok2010}). 
The spin period of PSR J1906+0746 is 0.144s, and it is likely that the $\rho \propto P_{\rm s}^{-0.5}$ relation works with this pulsar (see fig.\ 2 in O'Shaughnessy \& Kim (2010\nocite{ok2010}). 
More timing and long-term monitoring observations are needed, in order to pin down the pulsar's beaming correction factor.

PSR J1756$-$5521 is also not included in this work.
\citet{kkl2010} and \citet{ok2010} calculated \prate~for this pulsar.
It is arguably the most uncertain among the known NS$-$NS binaries, 
because the selection effects for 
acceleration search that discovered this pulsar are only approximated in modelling. 
The contribution from PSR J1756$-$5521 is expected to be 
roughly a few per cent in ${\cal R}_{\rm g}$ \citep{kkl2010} and
is comparable to that of PSR B1534+12 (see fig.\ 7 in \citet{ok2010}),
if its beam function follows the empirical $\rho$-$\pspin$ relation.

The Galactic NS$-$NS merger rate estimated in this work is based on properties of known NS$-$NS mergers at the current epoch. 
The same approach is applied in previous works on the empirical NS$-$NS merger rate estimates such as \citet{kalogera2004} and \citet{ok2010} as well. 
In time, the Double Pulsar will consist of the A pulsar and a radio-quiet NS after B’s become radio-quiet. Assuming continuous formation and merger of NS$-$NS binaries over the age of our Galaxy, it is possible that there are binaries consisting of a radio-active A and a radio-quiet B. These binaries must be older than B’s radio emission time-scale and would be expected to have tighter, more circular orbits. Moreover, the detectable pulsar in the binary would spin more slowly than A. In order to preform survey simulations for these binaries, different observational biases are required. Monte Carlo simulations of such systems with no detection require free parameters, most importantly the epoch of detection. Although it is technically possible to model a few different orbital configurations assuming binary orbital evolution (e.g. Peters \& Mathews 1963) and simple spin-down for A, the uncertainties involved would substantially increases uncertainties in our rate estimates. In this work, we therefore calculate the Galactic NS$-$NS merger rate estimates only using the best observational constraints available at present, especially for the A and B pulsars.

       \begin{center}
       \begin{figure} 
          \includegraphics[width=8.5cm]{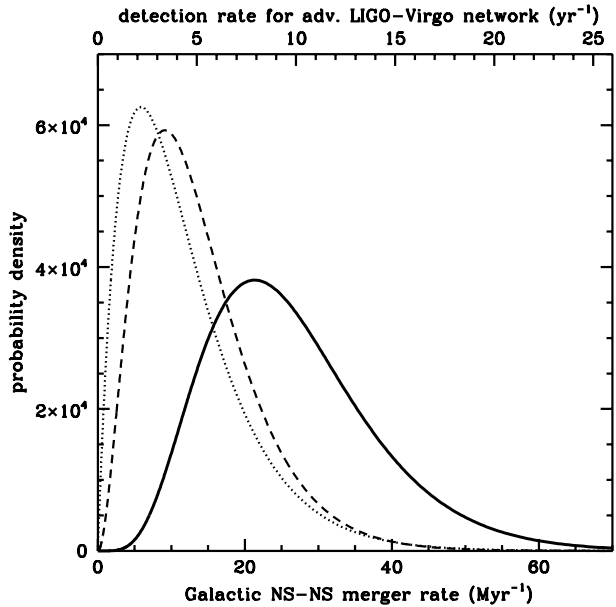}
          \caption{${\cal P}_{\rm g}$(${\cal R}_{\rm g}$)
                   (solid) is overlaid with individual \prate~obtained from 
                   PSR B1916+13 (dotted) and the Double Pulsar (short dashed). 
                   Based on our reference model,
                   the Galactic NS$-$NS merger rate is most likely to be $21$ Myr$^{-1}$. 
                   The corresponding GW detection rate 
                   for the advanced ground-based GW detectors is $\sim 8$ yr$^{-1}$.}
           \label{fig:ratesall}
       \end{figure}
       \end{center}

In order to better constrain the contribution of known pulsar binaries
to the Galactic NS$-$NS merger rate estimates, 
we call for a more realistic surface magnetic field 
and/or radio emission model.
A binary formation model that can describe the spin evolution of 
A and B (to pin down the binary age) is also useful.
Additional pulse profile observations of B will be 
invaluable to map out its beam function more accurately when it reappears.

More discoveries of relativistic NS$-$NS binaries are also important.
Large-scale pulsar surveys with unprecedented sensitivity such as
the LOFAR (LOw Frequency ARray; van Leeuwen \& Stappers 2010\nocite{lofarpulsar2010}) and the planned 
Square Kilometre Array \citep{smits2009ska} 
are expected to find more NS$-$NS binaries.
In addition to electromagnetic wave surveys, GW detection 
will provide a completely new, independent probe for 
relativistic NS$-$NS binaries. When the ground-based GW detectors 
start detecting NS$-$NS binaries or pulsar-black hole binaries, 
those {\it observed} GW detection rate will be useful  
to further constrain the pulsar population models.

\section*{Acknowledgements}

The authors are grateful for referee's comments.
MAM and CK were supported by a WVEPSCoR Research Challenge Grant. 
CK is supported in part by the National Research Foundation Grant 
funded by the Korean Government (no.\ NRF-2011-220-C00029). 
BP is supported by a NRAO student support award. 
The authors thank W.\ M.\ Farr, D.\ R.\ Lorimer, and R.\ O'Shaughnessy 
for critical reading of the manuscript. CK also thanks C.\ M.\ Miller and I.\ Mandel for useful discussions.

\end{document}